\documentstyle[aps,pra,epsf]{revtex}

\begin{document}

\title{Superfluid pairing in a polarized dipolar Fermi gas}

\author{M.A. Baranov$^{1,2,3}$, M.S. Mar'enko$^{5,6}$, Val.S. Rychkov$^{1,3}$ 
and G.V. Shlyapnikov$^{1,3,4}$}

\address{
(1) Russian Research Center Kurchatov Institute, Kurchatov Square,123182
Moscow, Russia\\
(2) Institut f\"ur Theoretische Physik, Universit\"at Hannover,
D-30167 Hannover, Germany\\
(3) FOM Institute for Atomic and Molecular Physics, Kruislaan 407, 1098 SJ
Amsterdam, The Netherlands\\
(4) Laboratoire Kastler Brossel\footnote{ 
LKB is an unit\'{e} de recherche de l'Ecole Normale 
Sup\'{e}rieure et de l'Universit\'{e} Pierre et Marie Curie, associ\'{e}e 
au CNRS.}, 
24 rue Lhomond, F-75231 Paris Cedex 05, France\\
(5) P.L. Kapitza Institute for Physical Problems, 117334 Moscow, Russia\\
(6) Institute of Radio Engineering and Electronics RAS, 103907 Moscow,
Russia
}

\maketitle
\vspace{1mm}
\begin{abstract}

We calculate the critical temperature of a superfluid phase transition in a
polarized Fermi gas of dipolar particles. In this case the order parameter
is anisotropic and has a nontrivial energy dependence. Cooper pairs do
not
 have a definite value of the angular momentum and are coherent
superpositions of all odd angular momenta. Our results describe prospects
for achieving the superfluid transition in single-component gases of
fermionic
 polar molecules.
\end{abstract}

\section{\protect\bigskip Introduction}

The recent success in observing quantum degeneracy in ultra-cold atomic
Fermi gases \cite{coolfer1,coolfer2,coolfer3,coolfer4} stimulates a search
for gaseous Fermi systems with an achievable temperature of superfluid phase
transition. The ideas based on Cooper pairing for a short-range Van der
Waals interaction between atoms \cite{coolgasBCS} require a simultaneous
trapping of at least two different fermionic species, with a rather severe
constraint on their relative concentrations. The situation is different for
Fermi gases of dipolar particles. Being electrically polarized, these
particles interact via long-range anisotropic (partially attractive)
dipole-dipole forces. In the ultra-cold limit, the dipole-dipole scattering
amplitude is energy independent for any orbital angular momenta \cite
{orbital}. This opens prospects to achieve the superfluid pairing in a {\it %
single-component} Fermi gas, where only scattering with odd orbital momenta 
(negligible in the case of Van der Waals interactions) is present. These
prospects are especially interesting as in single-component fermionic gases
the Pauli exclusion principle provides a strong suppression of inelastic
collisional rates (see \cite{Verhaar}). Hence, one can think of achieving
significantly higher densities than in Bose gases.

Possible realizations of dipolar Fermi gases include an electrically
polarized gas of polar molecules as they have large permanent  electric
dipoles. The creation of cold clouds of polar molecules has  been  recently
demonstrated in experiments with buffer-gas cooling \cite{buffercool} and in
experiments based on deceleration and cooling  of polar molecules by
time-dependent electric fields \cite{movefield}.  Another option is to
create a gas of atoms with electric dipole moments  induced by a high dc
electric field \cite{you-marinescu0} or by laser  coupling of the  atomic
ground state to an electrically polarized Rydberg  state \cite{rydberg}.

The $p$-wave Cooper pairing in a polarized dipolar Fermi gas has been
discussed in \cite{dipstoof,you-marinescu}, and the corresponding critical
temperature has been estimated by using the standard BCS approach. In this
paper we calculate the exact value of the critical temperature and find the
energy and angular dependence of the order parameter. For this purpose we
consider the Cooper pairing for all possible scattering channels. These
channels are coupled to each other by the dipole-dipole interaction, and the
Cooper pairs prove to be coherent superpositions of contributions of all odd
angular momenta. In order to find the pre-exponential factor for the critical
temperature, we perform the calculations to second order in perturbation
theory along the lines of the approach of Gor'kov and Melik-Barkhudarov (GM
approach) \cite{GM}.

\section{\protect\bigskip General equations}

We consider a spatially homogeneous single-component gas of fermions having
a dipole moment ${\bf d}$ oriented along the $z$-axis. The Hamiltonian of
the system has the form

\begin{equation}
H=\int d{\bf r}\widehat{\psi }^{\dagger }({\bf r})\left\{ -\frac{\hbar ^{2}}{%
2m}\nabla ^{2}-\mu \right\} \widehat{\psi }({\bf r})+\frac{1}{2}\int d{\bf r}%
d{\bf r}^{\prime }\left| \widehat{\psi }({\bf r})\right| ^{2}V_{d}({\bf r}-%
{\bf r}^{\prime })\left| \widehat{\psi }({\bf r}^{\prime })\right| ^{2},
\label{1}
\end{equation}
where $\widehat{\psi }({\bf r})$ is the field operator for fermions, $V_{d}(%
{\bf r})=(d^{2}/r^{3})(1-3\cos ^{2}\theta _{{\bf r}})$ the dipole-dipole
interaction, $\theta _{{\bf r}}$ is the angle between the interparticle
distance ${\bf r}$ and the $z$-axis, and $\mu $ is the chemical potential.
In Eq.(\ref{1}) we omit the contribution of the $p$-wave scattering due to
the short-range part of interparticle interaction, since this contribution is
small in the dilute ultra-cold limit.

For a single-component Fermi gas, the Cooper pairing is possible only in the
states with an odd angular momentum $l$. On the other hand, the anisotropic
character of the dipole-dipole interaction leads to coupling between Cooper
pairs with different values of the angular momentum. Therefore, the problem
of superfluid pairing requires us to consider states with any odd $l$.

The critical temperature $T_{c}$ of the superfluid transition and the order
parameter $%
 \Delta $ can be found from the gap equation in the momentum
representation 
 \cite{BCS,leggett}:

\begin{equation}
\Delta ({\bf p})=-\int \frac{d{\bf p}^{\prime }}{(2\pi \hbar )^{3}}V({\bf p},%
{\bf p}^{\prime })\frac{\tanh (E({\bf p}^{\prime })/2T)}{2E({\bf p}^{\prime
})}\Delta ({\bf p}^{\prime }).  \label{1.1}
\end{equation}
Here $E({\bf p})=\sqrt{\Delta ^{2}({\bf p})+(p^{2}/2m-\mu )^{2}}$, and we
assume the order parameter to be real. The function $V({\bf p},{\bf p}%
^{\prime })=V_{d}({\bf p-p}^{\prime })+\delta V({\bf p},{\bf p}^{\prime })$,
where $V_{d}({\bf q})$ is the Fourier transform of the dipole-dipole
interparticle interaction potential $V_{d}({\bf r})$:

\begin{equation}
V_{d}({\bf q})=\frac{4\pi }{3}d^{2}(3\cos ^{2}(\theta _{{\bf q}})-1),
\label{2'}
\end{equation}
with $\theta _{{\bf q}}$ being the angle between the momentum ${\bf q}$ and
the $z$-axis. The quantity $\delta V({\bf p},{\bf p}^{\prime })$ originates
from many-body effects and is a correction to the bare interparticle
interaction $V_{d}$. The leading corrections are second order in
$V_{d}$ and the
 corresponding diagrams are shown in Fig.1 (see Ref.
\cite{GM}). They describe the processes in which one of the two colliding
particles polarizes the medium by virtually creating a particle-hole pair. In
1a the particle-hole pair then annihilates due to the interaction with the
other colliding particle. In 1b, 1c and 1d the hole annihilates together with
one of the colliding particles. In 1b and 1c the particle-hole pair is
created due to the interaction of the medium with one of the colliding
particles, and the hole annihilates with the other colliding partner. In 1d
these  creation and annihilation processes involve one and the same colliding 
particle. 
\begin{figure}[tbp]
\epsfxsize10cm 
\centerline{
\epsfbox{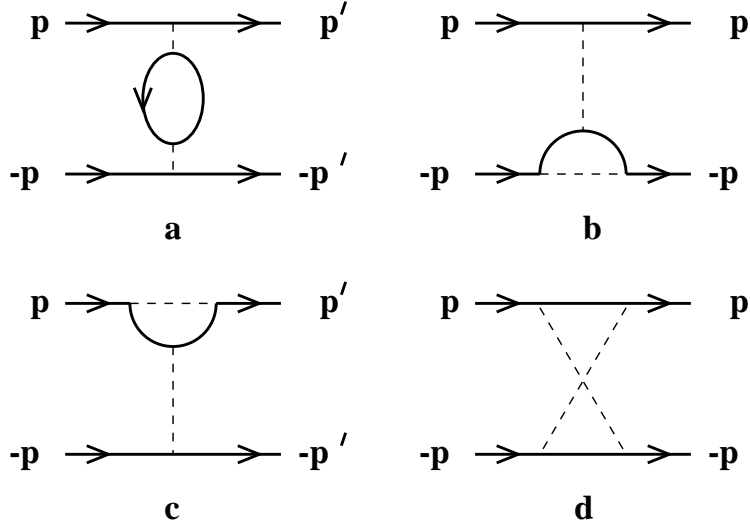}}
\caption{The lowest order many-body corrections to the effective interparticle
interaction.}
\end{figure}

For temperatures just below $T_{c}$ the order parameter is small and the
gap equation is equivalent to the Ginzburg-Landau equation for the spatially
homogeneous order parameter. This equation can be obtained by expanding the
rhs of Eq.(\ref{1.1}) in powers of the order parameter $\Delta ({\bf p})$:

\begin{equation}
\Delta ({\bf p})=-\int \frac{d{\bf p}^{\prime }}{(2\pi \hbar )^{3}}V({\bf p},%
{\bf p}^{\prime })\left[ K(p^{\prime })\Delta ({\bf p}^{\prime })+\,\frac{%
\partial K(p^{\prime })}{\partial \xi ^{\prime }}\frac{\Delta ^{3}({\bf p}%
^{\prime })}{2\xi ^{\prime }}\right] ,  \label{2}
\end{equation}
where $K(p)=\tanh (\xi /2T)/2\xi $, and $\xi =p^{2}/2m-\mu $.

The occurrence of the Cooper pairing is associated with the existence of a
nontrivial solution of Eq.(\ref{2}) for temperatures $T\leq T_{c}$. In order
to find the value of the critical temperature one can neglect the second,
nonlinear term in the square brackets in the rhs of Eq.(\ref{2}) because
for $T\rightarrow T_{c}$ the order parameter $\Delta \rightarrow 0$. The
corresponding linearized gap equation also provides us with the momentum
dependence of the order parameter, whereas the nonlinear term determines the
absolute (temperature dependent) value of $\Delta $.

The integral in Eq.(\ref{2}) diverges at large momenta. The divergency can
be eliminated by expressing the bare interaction $V_{d}$ in terms of the
vertex function ( scattering off-shell amplitude) $\Gamma (E,{\bf p},{\bf p}%
^{\prime })$. This is similar to the well-known procedure of renormalization
of the scattering length in dilute gases of Bose or Fermi particles
interacting via short-range forces \cite{LL9,AGD}. One may choose any value of
$%
 E $, and for simplifying our calculations we select $E=0$. Then the vertex
function $\Gamma (0,{\bf p},{\bf p}^{\prime })=\Gamma _{d}({\bf p},{\bf p}%
^{\prime })$ obeys the equation

\begin{equation}
\Gamma _{d}({\bf p},{\bf p}^{\prime })=V_{d}({\bf p-p}^{\prime })-\int \frac{%
d{\bf q}}{(2\pi \hbar )^{3}}\Gamma _{d}({\bf p},{\bf q})K_{0}(q)V_{d}({\bf %
q-p}^{\prime }),  \label{3}
\end{equation}
with $K_{0}(q)=m/q^{2}$. We will confine ourselves to the second order in
perturbation theory with respect to $V_{d}$. Omitting higher order
corrections, the renormalized linearized gap equation reads

\begin{equation}
\Delta ({\bf p})=-\int \frac{d{\bf p}^{\prime }}{(2\pi \hbar )^{3}}\Gamma
_{d}({\bf p,p}^{\prime })\left\{ K(p^{\prime })-K_{0}(p^{\prime })\right\}
\Delta ({\bf p}^{\prime })-\int \frac{d{\bf p}^{\prime }}{(2\pi \hbar )^{3}}
\delta V({\bf p},{\bf p}^{\prime })K(p^{\prime })\Delta ({\bf p}^{\prime }).
\label{4}
\end{equation}

In the dilute ultra-cold limit only small momenta ${\bf p}$ and ${\bf p}
^{\prime }$ are important. We thus have to find the scattering amplitude for
ultra-cold particles, 
in the presence of the dipole-dipole interaction between them.

\section{\protect\bigskip Scattering amplitude in the ultra-cold limit}

The anisotropic and long-range character of the dipole-dipole interaction ($%
V_d\propto 1/r^3$) ensures that in the ultra-cold limit all partial waves
give an energy independent contribution to the scattering amplitude \cite
{orbital}.  For any orbital angular momentum $l$ one has $\Gamma_d\sim
d^2\sim 4\pi\hbar^2r_*/m$, where the quantity $r_*\sim md^2/\hbar^2$ plays a
role of the characteristic radius of interaction for the dipole-dipole
potential. For the interparticle separation $r\gg r_*$ the potential $V_d(%
{\bf r})$ does not influence the wave function of the relative motion of two
colliding particles and the motion becomes free. The ultra-cold limit requires
particle momenta satisfying the condition 
\begin{equation}  \label{limit}
pr_*/\hbar\ll 1.
\end{equation}

The anisotropy of $V_d$ directly couples scattering channels with angular
momenta $l$ and $l\pm 2$. Thus, strictly speaking, all even-$l$ (odd-$l$)
channels are coupled to each other, whereas the scattering with odd angular
momenta remains decoupled from that with even momenta.

There are two contributions to the scattering amplitude. The long-range
contribution comes from distances $r\agt \hbar/p$ and gives $\Gamma_d\sim d^2
$ for all angular momenta in the incoming and outgoing channels, allowed by
the selection rules. This contribution can be found by using the Born
approximation. The short-range contribution comes from distances $r\alt r_*$%
. For the scattering with even $l$, due to the presence of the $s$-wave, we
have again $\Gamma_d\sim d^2$ or even somewhat larger because of the
so-called shape resonances \cite{You}. Under the condition (\ref{limit}),
the contribution of the $s$-wave to the wave function of the relative motion
at distances $r\alt r_*$ is independent of $p$. This leads to an
energy independent $\Gamma$.  However, it depends on a detailed behavior of the
interparticle potential at short interparticle distances. Thus, for even $l$
one can not make a general statement on
the value of $\Gamma$.

In the case of identical fermions only odd orbital angular momenta are
present. Then the short-range contribution is much smaller than the
long-range one. We will demonstrate this for the $p$-wave on-shell
scattering amplitude, omitting the coupling to the channels with other odd $l
$. For $l=1$ and $m_l=0$ in both incoming and outgoing scattering channels ($%
m_l$ is the projection of $l$ on the $z$ axis), the dipole-dipole potential $%
V_d({\bf r})$ averaged over the angle $\theta_{{\bf r}}$ is equal to ${%
\bar V}_d=-4d^2/5r^3$. In order to analyze the short- and long-range
contributions to the scattering amplitude, we consider the relative motion
of particles in a truncated potential $V(r)={\bar V}_d(r)$ for $r<r_0$, 
and $V(r)=0$ for $r>r_0$.

The truncation radius $r_0$ is selected such that $r_*\alt r_0\ll \hbar/p$.
The Schr\"odinger equation for the wave function of the relative motion
reads 
\begin{equation}  \label{Schr}
\frac{\hbar^2}{m}\left(-\frac{d^2}{dr^2}-\frac{2}{r}\frac{d}{dr}+ \frac{6}{%
r^2}\right)\psi(r) + V(r)\psi(r)=\frac{p^2}{m}\psi(r).
\end{equation}

At distances $r\ll \hbar /p$ we may put $p=0$ in Eq.(\ref{Schr}). Then for 
$r<r_{0}$ we use the well-known procedure of finding an analytical solution
for $\psi (r)$ in power law potentials \cite{LL3}. Assuming $r\gg r_{*}$,
this gives 
\begin{equation}
\psi (r)\propto \left( \frac{r}{r_{*}}+{\rm const}\right) ,  \label{psi1}
\end{equation}
where the constant term is independent of $r_{*}$. At $r>r_{0}$ the motion
is free and $\psi (r)$ depends explicitly on the scattering phase $\delta$. 
The solution, which for $r\rightarrow \infty $ takes the required
asymptotic form $(\hbar /pr)\cos {(pr/\hbar +\delta )}$, at $r\ll \hbar /p$
becomes 
\begin{equation}
\psi (r)=-\left( \frac{\sin \delta }{(pr/\hbar )^{2}}+\frac{(pr/\hbar )\cos
\delta }{3}\right) .  \label{psi2}
\end{equation}
Equalizing the logarithmic derivatives of the wave functions (\ref{psi1})
and (\ref{psi2}), we immediately obtain $\delta \sim
p^{3}r_{0}^{2}r_{*}/\hbar ^{3}$ and find that $\psi (r)\propto p$ for $r\ll
\hbar /p$. The scattering amplitude then proves to be $\Gamma _{d}=\int \psi
(r)V(r)d^{3}r\sim d^{2}(pr_{0}/\hbar )^{2}$. The short-range contribution to 
$\Gamma $, that is the contribution from distances $r\alt r_{*}$, is
obtained from this relation by simply putting $r_{0}\sim r_{*}$.

We then increase $r_0$ and make it much larger than $\hbar/p$. At distances $%
r\sim \hbar/p\gg r_*$ the potential ${\bar V}$ is much smaller than the
kinetic energy term in the lhs of Eq.(\ref{Schr}). For the contribution
of these distances to the scattering amplitude the Born approximation gives $%
\Gamma_d\sim d^2$. We thus see that the short-range contribution to the
scattering amplitude is small as $(pr_*/\hbar)^2$ compared to the long-range
contribution coming from distances $r\sim\hbar/p$.

This has two important consequences. First, a detailed shape of the
interaction potential is not important for the scattering amplitude as the
latter is determined by the long-range contribution. This contribution is
obtained in the Born approximation and depends only on the value of the
dipole moment. Second, we may include the second order Born correction to
the amplitude. This correction is of the order of $d^2(pr_*/\hbar)$ and
still greatly exceeds the short-range contribution.

In the second order Born approximation for the off-shell scattering
amplitude $\Gamma _{d}({\bf p},{\bf p}^{\prime })$ we have 
\begin{equation}
\Gamma _{d}({\bf p},{\bf p}^{\prime })=V_{d}({\bf p-p}^{\prime })-\int \frac{%
d{\bf q}}{(2\pi \hbar )^{3}}V_{d}({\bf p-q})K_{0}(q)V_{d}({\bf q-p}^{\prime
}),  \label{6}
\end{equation}
where the first and second terms in the rhs of Eq.(\ref{6}) are first and
second order in $V_{d}$, respectively. The integral for the second order
correction to the scattering amplitude in Eq.(\ref{6}) is formally divergent at
large $q$. This is the same non-physical divergency as in the case of
short-range interactions \cite{LL9,AGD}, and it will be eliminated in the
calculations of the order parameter and critical temperature (see Section V).  

\section{\protect\bigskip Critical temperature in the BCS approach}

In a quantum degenerate Fermi gas characteristic momenta of colliding
particles are of the order of the Fermi momentum $p_F=(6\pi ^{2}n)^{1/3}
\hbar $ ($n$ is the gas density). Then, with $r_*\sim md^2/\hbar^2$, the
condition (\ref{limit}) of the ultra-cold limit for interparticle collisions
can be written as 
\begin{equation} 
nd^{2}/\varepsilon _{F} \ll 1,  \label{sparameter} 
\end{equation} 
where $\varepsilon_F=p_F^2/2m$ is the Fermi energy. The lhs of
Eq.(\ref{sparameter}) is the ratio of the mean dipole-dipole interaction
energy (per particle) to the Fermi energy. As in the case of
short-range interactions \cite{LL9,AGD}, this is a small parameter of the
many-body theory. It is the condition (\ref{sparameter}) that allows us to
omit the contribution of higher order diagrams and use the renormalized gap
equation (\ref{4}).      

Generally, in dilute Fermi gases the critical temperature is exponentially
small compared to the Fermi energy $\varepsilon _{F}$. The exponent is
inversely proportional to the Fermi momentum $p_{F}$ and is determined by 
first order terms in $V_{d}$. The account of the second order
terms provides us with the pre-exponential factor.

We first calculate $\Delta ({\bf p})$ to first order in $V_{d}$ and find the
correct exponent in the dependence of the critical temperature on the
particle density. For this purpose we should keep in Eq.(\ref{4}) only the
terms which are first order in $V_{d}$. This is the first term in the rhs of
this equation, with $\Gamma _{d}({\bf p},{\bf p}^{\prime })=V_{d}({\bf p-p}%
^{\prime })$. Then, Eq.(\ref{4}) can be rewritten in the form
\begin{equation}
\Delta (\xi ,{\bf n})=-\int_{-\mu }^{\infty }d\xi ^{\prime }(\tanh (\xi
^{\prime }/2T)/2\xi ^{\prime })\int \frac{d{\bf n}^{\prime }}{4\pi }R(\xi ,%
{\bf n};\xi ^{\prime },{\bf n}^{\prime })\Delta (\xi ^{\prime },{\bf n}%
^{\prime }).  \label{5}
\end{equation}
Here \ ${\bf n}={\bf p}/p$, and
\[
R(\xi ,{\bf n};\xi ^{\prime },{\bf n}^{\prime })=\nu (\xi ^{\prime })\Gamma
_{d}(p(\xi ){\bf n},p(\xi ^{\prime }){\bf n}^{\prime })\left( 1-\xi ^{\prime
}/(\xi ^{\prime }+\mu )\tanh (\xi ^{\prime }/2T)\right) , 
\]
where $\nu (\xi )=mp(\xi )/2\pi ^{2}\hbar ^{3}$ is the density of states at
energy $\xi +\mu $. The chemical potential $\mu $ is equal to the Fermi
energy: $\mu =\varepsilon _{F}$.

The main contribution to the pairing comes from the states near the Fermi
surface, where $\left| \xi \right| ,\left| \xi ^{\prime }\right| \ll
\varepsilon _{F}$. In order to single out this contribution in the rhs of
Eq.(\ref{5}), we introduce a characteristic energy $\overline{\omega }$
that obeys the constraint $T\ll \overline{\omega }$, and is of the order of
the Fermi energy. We then divide the integral over $\xi ^{\prime }$ in Eq.(%
\ref{5}) into two parts: (a) the integration of $R(\xi ,{\bf n};0,{\bf n}%
^{\prime })\Delta (0,{\bf n}^{\prime })$ from $-\overline{\omega }$ to $%
\overline{\omega }$, and (b) the integration of $\left[ R(\xi ,{\bf n};\xi
^{\prime },{\bf n}^{\prime })\Delta (\xi ^{\prime },{\bf n}^{\prime })-R(\xi
,{\bf n};0,{\bf n}^{\prime })\Delta (0,{\bf n}^{\prime })\right] $ from $-%
\overline{\omega }$ to $\overline{\omega }$, plus the integration of $R(\xi ,%
{\bf n};\xi ^{\prime },{\bf n}^{\prime })\Delta (\xi ^{\prime },{\bf n}%
^{\prime })$ from $-\varepsilon _{F}$ to $-\overline{\omega }$ and from $%
\overline{\omega }$ to $\infty $. In part (a) we use the asymptotic formula
\[
\int_{-\overline{\omega }}^{\overline{\omega }}d\xi ^{\prime }(\tanh (\xi
^{\prime }/2T)/2\xi ^{\prime })\approx \ln \frac{2\exp (\gamma )\overline{%
\omega }}{\pi T}, 
\]
where $\gamma =0.5772$ is the Euler constant. In part (b) we replace $\tanh
(\xi ^{\prime }/2T)$ by the step function (omitting the unimportant
contribution from a narrow interval $\left| \xi ^{\prime }\right| \lesssim
T\ll \overline{\omega }$) and integrate in parts. As a result, Eq.(\ref{5})
takes the form
\begin{equation}
\Delta (\xi ,{\bf n})=-\ln \left[\frac{2\exp (\gamma )\overline{\omega }} {%
\pi T}\right]\int \frac{d{\bf n}^{\prime }}{4\pi } R(\xi ,{\bf n};0,{\bf n}%
^{\prime })\Delta (0,{\bf n}^{\prime }) +\frac{1}{2}\int_{-\varepsilon
_{F}}^{\infty}d\xi ^{\prime } \ln \frac{\left| \xi ^{\prime }\right| }{%
\overline{\omega}} \frac{d}{d\left| \xi ^{\prime }\right| } \int\frac{d{\bf n%
}^{\prime }}{4\pi } \left\{ R(\xi ,{\bf n};\xi ^{\prime},{\bf n}^{\prime })
\Delta(\xi^{\prime },{\bf n}^{\prime }) \right\} ,  \label{5'}
\end{equation}
where the first and second terms in the rhs come from parts (a) and (b),
respectively.

One can easily see that the second term in Eq.(\ref{5'}) is small as $1/\ln
(2\exp (\gamma )\overline{\omega }/\pi T)$ compared to the first one.
Therefore, the second term is only important for the pre-exponential factor
in the expression for the critical temperature and will be omitted in this
Section. This is equivalent to the commonly used BCS approach where the
kernel $R(\xi ,{\bf n};\xi ^{\prime },{\bf n}^{\prime })$ is replaced by $%
R(0,{\bf n};0,{\bf n}^{\prime })$ for $\left| \xi \right| ,\left| \xi
^{\prime }\right| \leq \overline{\omega }$ and by zero otherwise. Putting $%
\xi =0$ in Eq.(\ref{5'}), we obtain the following equation for finding the
critical temperature:
\begin{equation}
\Delta (0,{\bf n})=-\ln \left[\frac{2\exp (\gamma )\overline{\omega }}{\pi T}%
\right] \int \frac{d{\bf n}^{\prime }}{4\pi }R(0,{\bf n};0,{\bf n}^{\prime
}) \Delta (0,{\bf n}^{\prime }).  \label{5''}
\end{equation}

The anisotropic character of the scattering amplitude leads to a nontrivial
angular dependence of the order parameter $\Delta (0,{\bf n})$. In order to
analyze the possibility of pairing we expand $\Delta (0,{\bf n})$ in terms
of a complete set of eigenfunctions $\phi _{s}({\bf n})$ of the integral
operator with the kernel $R(0,{\bf n};0,{\bf n}^{\prime })$:

\begin{eqnarray}
\Delta (0,{\bf n)} &=&\sum_{s=0}^{\infty }\Delta _{s}\phi _{s}({\bf n}),
\label{7} \\
\int \frac{d{\bf n}^{\prime }}{4\pi }R(0,{\bf n};0,{\bf n}^{\prime })\phi
_{s}({\bf n}^{\prime }) &=&\lambda _{s}\phi _{s}({\bf n}),\quad s=0,1,\ldots
\label{7.1}
\end{eqnarray}
The functions $\phi _{s}({\bf n})$ are normalized by the condition $\int (d 
{\bf n}/4\pi )\phi _{s}^{2}({\bf n})=1$, and they are labeled by the index $%
s$ in such a way that the eigenvalues $\lambda_s<\lambda_{s+1}$.
Then Eq.(\ref{5''}) reduces to a set of equations

\[
\Delta _{s}\left( 1+\lambda _{s}\ln \frac{2\exp (\gamma )\overline{\omega }%
} {\pi T}\right) =0. 
\]

The appearance of a nontrivial solution for $\Delta (0,{\bf n)}$ below a
certain critical temperature requires the presence of at least one negative
eigenvalue $\lambda _{s}$. For a single eigenvalue $\lambda _{s^{\ast }}<0$,
the critical temperature immediately follows from the condition $(1+\lambda
_{s^{\ast }}\ln (2\exp (\gamma )\overline{\omega }/\pi T_{c}))=0$, and we
have $\Delta _{s^{\ast }}\neq 0$ and $\Delta _{s}=0$ for $s\neq s^{\ast }$.
In the case of several negative eigenvalues $\lambda _{s}<0$, one has to
choose the solution that corresponds to the lowest eigenvalue as it gives
the highest critical temperature.

Using Eq.(\ref{2'}) one finds that negative $\lambda _{s}$ correspond to
eigenfunctions $\phi _{s}$ which are independent of the azimuthal angle $%
\varphi$. This means that only spherical harmonics with zero projection $m$
of the angular momentum $l$ appear in their decomposition. For these
functions the kernel $R(0,{\bf n};0,{\bf n}^{\prime })$ \ can be reduced to
its average over the azimuthal angles $\varphi $ and $\varphi ^{\prime }$.
Using Eq.(\ref{2'}) for $\Gamma _{d}(p_{F}{\bf n},p_{F}{\bf n} ^{\prime })$,
we obtain
\begin{equation}
R(0,\cos \theta ;0,\cos \theta ^{\prime })=2\pi \frac{nd^{2}}{
\varepsilon _{F}}\left( \frac{3}{2}\left| \cos \theta -\cos \theta
^{\prime }\right| -1\right) ,  \label{7'}
\end{equation}
where $\theta $ and $\theta ^{\prime }$ are the polar angles for the
vectors ${\bf n}$ and ${\bf n}^{\prime }$, and $n$ is the gas density. Note
that the first multiple in the rhs of Eq.(\ref{7'}) is a small parameter
of the theory, given by Eq.(\ref{sparameter}) and representing the ratio of 
the mean dipole-dipole interaction energy to the Fermi energy $\varepsilon _{F}$. 

Keeping in mind that due to the Pauli principle only odd angular momenta are
present, we obtain the solutions of Eq.(\ref{7}):
\begin{equation}
\phi _{s}({\bf n})=\sqrt{2}\sin \left( \frac{\pi }{2}(1+2s)\cos (\theta
)\right) ,\ \lambda _{s}=-\frac{nd^{2}}{\varepsilon _{F}}\frac{12}{\pi
(1+2s)^{2}}.  \label{eigen}
\end{equation}
The lowest eigenvalue \ is $\lambda _{0}=-12nd^{2}/\pi \varepsilon _{F}$.
Therefore, the angular dependence of the order parameter will be
characterized by the function $\phi _{0}({\bf n})$ (see Section VI for
details). The critical temperature  is then given by
\begin{equation}
T_{c}=\frac{2\exp (\gamma )\overline{\omega }}{\pi }\exp \left( -\frac{1}{%
\left| \lambda _{0}\right| }\right) .  \label{9}
\end{equation}

In the BCS approach the pre-exponential factor ($\overline{\omega }$)
remains undetermined. One can only argue that it is of the order of $%
\varepsilon _{F}$. We thus have
\begin{equation}
T_{c}^{{\rm BCS}}\sim \varepsilon _{F}\exp \left( -\frac{\pi \varepsilon
_{F} }{12nd^{2}}\right) .  \label{TBCS}
\end{equation}
In Ref. \cite{you-marinescu} the exponent in the expression for $T_{c}^{{\rm %
BCS}}$ is only expressed in terms of the scattering amplitude which should
be found from the solution of a set of coupled equations. The estimate for
this exponent in Ref. \cite{dipstoof} takes into account only the $p-p$
scattering channel and contains a numerical error.

In order to find the pre-exponential factor one has to include the
contribution from the second term in Eq.(\ref{5'}), together with the
second order corrections to the eigenvalue $\lambda _{0}$. These corrections
originate from the second order many-body effects and from the second order
corrections to the scattering amplitude, described by the second terms in
Eqs.(\ref{4}) and (\ref{6}) respectively.

\section{\protect\bigskip GM approach. The calculation of the pre-exponential
factor}

We now proceed with the calculation of the pre-exponential factor in the
expression (\ref{9}) for the critical temperature. We first consider
the contribution of the second term in the rhs of Eq.(\ref{5'}), which is
logarithmically small compared to the already calculated first term. For
this purpose we specify the value of $\overline{\omega }$ by the condition
\begin{equation}
\int \frac{d{\bf n}}{4\pi }\phi _{0}({\bf n}) \int_{-\varepsilon
_{F}}^{\infty }d\xi ^{\prime } \ln \frac{\left| \xi ^{\prime }\right| }{%
\overline{\omega }} \frac{d}{d\left| \xi ^{\prime }\right| } \int \frac{d%
{\bf n}^{\prime }}{4\pi } \left\{ R(0,{\bf n};\xi ^{\prime },{\bf n}^{\prime
}) \Delta (\xi ^{\prime },{\bf n}^{\prime })\right\} =0.  \label{7''}
\end{equation}
Then, using Eqs.(\ref{7}) and (\ref{7.1}) we obtain the following
expression for $\overline{\omega }$:
\begin{equation}
\ln \overline{\omega }=-\frac{1}{\lambda _{0}} \int \frac{d{\bf n}}{4\pi }%
\phi _{0}({\bf n}) \frac{1}{2} \int_{-\varepsilon _{F}}^{\infty }d\xi
^{\prime } \ln \left| \xi ^{\prime}\right| \frac{d}{d\left| \xi ^{\prime
}\right| } \int \frac{d{\bf n}^{\prime }}{4\pi } \left\{ R(0,{\bf n}%
;\xi^{\prime },{\bf n}^{\prime }) \frac{\Delta (\xi ^{\prime },{\bf n}%
^{\prime })}{\Delta _{0}}\right\} .  \label{8'}
\end{equation}
This definition of $\overline{\omega }$ immediately leads to Eq.(\ref{9})
for the critical temperature and allows us to rewrite Eq.(\ref{5'}) in the
form
\[
\Delta (\xi ,{\bf n})=\frac{1}{\lambda _{0}}\int \frac{d{\bf n}^{\prime }}{%
4\pi }R(\xi ,{\bf n};0,{\bf n}^{\prime })\Delta (0,{\bf n}^{\prime }) 
\]
\begin{equation}
-\int \frac{d{\bf n}^{\prime }}{4\pi }\cdot \frac{1}{2}\int_{-\varepsilon
_{F}}^{\infty }\frac{d\xi ^{\prime }}{\left| \xi ^{\prime }\right| }\left\{
R(\xi ,{\bf n};\xi ^{\prime },{\bf n}^{\prime })\Delta (\xi ^{\prime },{\bf n%
}^{\prime })-\frac{R(\xi ,{\bf n};0,{\bf n}^{\prime })}{\lambda _{0}}\frac{%
\Delta (0,{\bf n}^{\prime })}{\Delta _{0}}\int \frac{d{\bf n}_{1}}{4\pi }%
\int \frac{d{\bf n}_{2}}{4\pi }\phi _{0}({\bf n}_{1})R(0,{\bf n}_{1};\xi
^{\prime },{\bf n}_{2})\Delta (\xi ^{\prime },{\bf n}_{2})\right\} ,
\label{9'}
\end{equation}
where the second term in the rhs is proportional to the small parameter of
the theory $nd^{2}/\varepsilon _{F}$ and can thus be considered as
a
perturbation. This follows from the fact that the bracket in this term
vanishes for $\xi ^{\prime }\rightarrow 0$. As a result, in contrast to the
first term of the rhs, the second term does not contain the large logarithm $%
\ln (\overline{\omega }/T)\sim \lambda _{0}^{-1}\sim (\varepsilon
_{F}/nd^{2})$.

The leading contribution to the angular dependence of the order parameter on
the Fermi surface comes from the term with $s=0$ in Eq.(\ref{7}): $\Delta
(0,{\bf n})=\Delta _{0}\phi _{0}({\bf n})$. Therefore, to the leading order
in $nd^{2}/\varepsilon _{F}$, the solution of Eq.(\ref{9'}) is
\begin{equation}
\Delta (\xi ,{\bf n})\approx \frac{1}{\lambda _{0}}\int \frac{d{\bf n}%
^{\prime }}{4\pi }R(\xi ,{\bf n};0,{\bf n}^{\prime })\Delta _{0}\phi _{0}(%
{\bf n}^{\prime }).  \label{8}
\end{equation}
After substituting this expression into Eq.(\ref{8'}) and performing a
numerical integration, we obtain
\begin{equation}
\overline{\omega }\approx \exp \left( -\frac{1}{\lambda _{0}} \int \frac{d%
{\bf n}}{4\pi }\phi _{0}({\bf n}) \frac{1}{2}\int_{-\varepsilon
_{F}}^{\infty }d\xi ^{\prime } \ln \left| \xi^{\prime }\right| \frac{d}{%
d\left| \xi ^{\prime }\right| } \int \frac{d{\bf n}^{\prime }}{4\pi }
\left\{ R(0,{\bf n};\xi ^{\prime },{\bf n}^{\prime }) \frac{\Delta (\xi
^{\prime },{\bf n}^{\prime })}{\Delta _{0}}\right\} \right) =0.42\varepsilon
_{F},  \label{10}
\end{equation}
Corrections to Eqs.(\ref{8}) and (\ref{10}) are related to the terms with $%
s\neq 0$ in Eq.(\ref{7}), and from Eq.(\ref{9'}) we find that the
quantities $\Delta _{s\neq 0}\sim \Delta _{0}(nd^{2}/\varepsilon _{F})$.
These corrections lead to the relative contribution of the order of $%
nd^{2}/\varepsilon _{F}$ to the pre-exponential factor for the critical
temperature and hence will be neglected.

We now calculate the contributions from the second terms in Eqs.(\ref{4})
and (\ref{6}). As one can see from Eq.(\ref{7}), these terms result in the
correction for the eigenvalue $\lambda _{0}$:
\begin{equation}
\delta \lambda _{0}=\nu (0)\int \frac{d{\bf n}}{4\pi }\int \frac{d{\bf n}%
^{\prime }}{4\pi }\phi _{0}({\bf n})\left\{ \delta V(p_{F}{\bf n},p_{F}{\bf n%
}^{\prime })-\int \frac{d{\bf q}}{(2\pi \hbar )^{3}}V_{d}({\bf p-q}%
)K_{0}(q)V_{d}({\bf q-p}^{\prime })\right\} \phi _{0}({\bf n}^{\prime }).
\label{10.1}
\end{equation}
The first term in the integrand of Eq.(\ref{10.1}) originates from
many-body effects, and the quantity\ $\delta V({\bf p},{\bf p}^{\prime })$
is shown diagrammatically in Fig.1. The analytical expressions for the
diagrams in Figures from 1a to 1d, read:
\begin{eqnarray*}
\delta V_{a}({\bf p},{\bf p}^{\prime }) &=&\int \frac{d{\bf q}}{(2\pi )^{3}}%
\frac{N({\bf q}+{\bf p}_{-}/2)-N({\bf q}-{\bf p}_{-}/2)}{\xi _{{\bf q}+{\bf p%
}_{-}/2}-\xi _{{\bf q}-{\bf p}_{-}/2}}V_{d}^{2}({\bf p}_{-}), \\
\delta V_{b}({\bf p},{\bf p}^{\prime }) &=&-\int \frac{d{\bf q}}{(2\pi )^{3}}%
\frac{N({\bf q}+{\bf p}_{-}/2)-N({\bf q}-{\bf p}_{-}/2)}{\xi _{{\bf q}+{\bf p%
}_{-}/2}-\xi _{{\bf q}-{\bf p}_{-}/2}}V_{d}({\bf p}_{-})V_{d}({\bf q}-{\bf p}%
_{+}/2), \\
\delta V_{c}({\bf p},{\bf p}^{\prime }) &=&-\int \frac{d{\bf q}}{(2\pi )^{3}}%
\frac{N({\bf q}+{\bf p}_{-}/2)-N({\bf q}-{\bf p}_{-}/2)}{\xi _{{\bf q}+{\bf p%
}_{-}/2}-\xi _{{\bf q}-{\bf p}_{-}/2}}V_{d}({\bf p}_{-})V_{d}({\bf q}+{\bf p}%
_{+}/2), \\
\delta V_{d}({\bf p},{\bf p}^{\prime }) &=&-\int \frac{d{\bf q}}{(2\pi )^{3}}%
\frac{N({\bf q}+{\bf p}_{+}/2)-N({\bf q}-{\bf p}_{+}/2)}{\xi _{{\bf q}+{\bf p%
}_{+}/2}-\xi _{{\bf q}-{\bf p}_{+}/2}}V_{d}({\bf q}-{\bf p}_{-}/2)V_{d}({\bf %
q}+{\bf p}_{-}/2).
\end{eqnarray*}
Here ${\bf p}_{\pm }={\bf p}\pm {\bf p}^{\prime }$, and $N({\bf p})$ is the
Fermi-Dirac distribution at zero temperature. The integrals related to the
first term in the rhs of Eq.(\ref{10.1}), with $\delta V({\bf p},{\bf p}%
^{\prime })=\sum_{\alpha =a,...,d}\delta V_{\alpha }({\bf p},{\bf p}^{\prime
})$, were calculated numerically by using the Monte-Carlo method. Each of
the terms $\delta V_{\alpha }({\bf p},{\bf p}^{\prime })$ provides a
correction $\delta \lambda _{0}^{(\alpha )}=\left[ \nu (0)4\pi
d^{2}/3\right] ^{2}\eta _{\alpha }$. For the coefficients $\eta _{\alpha }$
we find $\eta _{a}=0.19$, $\eta _{b}=\eta _{c}=-0.08$, and $\eta _{d}=0.42$.
Thus, the first term in the rhs of Eq.(\ref{10.1}) gives the correction
\[
\delta \lambda _{0}^{(1)}=0.45\left( \nu (0)\frac{4\pi d^{2}}{3}\right)
^{2}. 
\]
\qquad

The second term in the integrand of Eq.(\ref{10.1}) comes from the second
order correction to the scattering amplitude $\Gamma _{d}$. For the
correction to $\lambda _{0}$, originating from this term, our numerical
calculation gives
\[
\delta \lambda _{0}^{(2)}=-0.86\left( \nu (0)\frac{4\pi d^{2}}{3}\right)
^{2}. 
\]
Note that the function $\phi_0({\bf n})$ is odd with respect to
$\cos{\theta}$. For this reason, the integration over $d{\bf n}$ and $d{\bf
n}'$ eliminates the formal divergency of the integral over $d{\bf q}$ at large
$q$. 

The total correction to the eigenvalue $\lambda _{0}$ is then
\begin{equation}
\delta \lambda _{0}=\delta \lambda _{0}^{(1)}+\delta \lambda
_{0}^{(2)}=-0.41\left( \nu (0)\frac{4\pi d^{2}}{3}\right) ^{2}.  \label{10.2}
\end{equation}
On the basis of Eqs.(\ref{9}), (\ref{10}) and (\ref{10.2}), we obtain
the final expression for the critical temperature:
\begin{equation}
T_{c}=\frac{2{\rm \exp }(\gamma )}{\pi }\times 0.42\,\varepsilon _{F}\exp
(-1/\left| \lambda _{0}+\delta \lambda _{0}\right| )\approx
1.44\,\varepsilon _{F}\exp (-\pi \varepsilon _{F}/12nd^{2}).  \label{11}
\end{equation}
It is worth noting that if we include only the $p-p$ scattering channel
the exponent in Eq.(\ref{11}) will be larger by a factor of $10/\pi^2$. The
pre-exponential factor becomes then larger by a factor of $1.1$.


\section{Anisotropic order parameter}

In order to find the temperature dependence of the order parameter for $%
T\leq T_{c}$, we have to include the nonlinear term in the gap equation (\ref
{5''}). This term can be written as
\[
\int \frac{d{\bf p}^{\prime }}{(2\pi \hbar )^{3}}V_{d}({\bf p-p}^{\prime
})\left[ \frac{1}{\cosh ^{2}(\xi ^{\prime }/2T)}-\frac{\tanh (\xi ^{\prime
}/2T)}{\xi ^{\prime }/2T}\right] \frac{\Delta ^{3}({\bf p}^{\prime })}{8\xi
^{\prime 2}T}, 
\]
where we neglect the many-body correction to the interparticle interaction.
The expression in the square brackets vanishes as $\left| \xi ^{\prime
}\right| ^{-3}$ for $\left| \xi ^{\prime }\right| \rightarrow \infty $.
Therefore, the main contribution to the integral comes from the region of
small $\xi ^{\prime }$, i.e. from ${\bf p}^{\prime }$ close to the Fermi
momentum $p_{F}$. This allows us to write

\begin{eqnarray*}
&&\int \frac{d{\bf p}^{\prime }}{(2\pi \hbar )^{3}}V_{d}({\bf p-p}^{\prime
})\left[ \frac{1}{\cosh ^{2}(\xi ^{\prime }/2T)}-\frac{\tanh (\xi ^{\prime
}/2T)}{\xi ^{\prime }/2T}\right] \frac{\Delta ^{3}({\bf p}^{\prime })}{8\xi
^{\prime 2}T} \\
&\approx &\int \frac{d{\bf n}^{\prime }}{4\pi }V_{d}({\bf p-}p_{F}{\bf n}%
^{\prime })\Delta ^{3}(p_{F}{\bf n}^{\prime })\int \frac{p^{\prime
2}dp^{\prime }}{2\pi ^{2}\hbar ^{3}}\left[ \frac{1}{\cosh ^{2}(\xi ^{\prime
}/2T)}-\frac{\tanh (\xi ^{\prime }/2T)}{\xi ^{\prime }/2T}\right] \frac{1}{%
8\xi ^{\prime 2}T} \\
&=&-\,\frac{7\zeta (3)}{8\pi ^{2}T^{2}}\int \frac{d{\bf n}^{\prime }}{4\pi }%
R(\xi ,{\bf n};0,{\bf n}^{\prime })\Delta ^{3}(0,{\bf n}^{\prime }),
\end{eqnarray*}
where $\zeta (z)$ is the Riemann zeta function. As a result, to first order
in $V_{d}$ the nonlinear gap equation reads:
\begin{equation}
\Delta (\xi ,{\bf n})=-\ln \frac{2\gamma \overline{\omega }}{\pi T}\int 
\frac{d{\bf n}^{\prime }}{4\pi }R(\xi ,{\bf n};0,{\bf n}^{\prime })\Delta (0,%
{\bf n}^{\prime })-\,\frac{7\zeta (3)}{8\pi ^{2}T^{2}}\int \frac{d{\bf n}%
^{\prime }}{4\pi }R(\xi ,{\bf n};0,{\bf n}^{\prime })\Delta ^{3}(0,{\bf n}%
^{\prime }).  \label{11.1}
\end{equation}

With the order parameter $\Delta (0,{\bf n})$ from Eq.(\ref{7}), where now $%
\Delta _{s}=\Delta _{s}(T)$ and $\Delta _{s}(T)\rightarrow 0$ for $%
T\rightarrow T_{c}$, Eq.(\ref{11.1}) takes the form
\begin{equation}
\sum_{s=0}^{\infty }\Delta _{s}\left( 1+\lambda _{s}\ln \frac{2\gamma 
\overline{\omega }}{\pi T}\right) \phi _{s}({\bf n})-\frac{7\zeta (3)}{8\pi
^{2}T^{2}}\sum_{s=0}^{\infty }\lambda _{s}\phi _{s}({\bf n})\left(
\sum_{\left\{ s_{i}\right\} }C_{s_{1}s_{2}s_{3}}^{s}\Delta _{s_{1}}\Delta
_{s_{2}}\Delta _{s_{3}}\right) =0.  \label{12}
\end{equation}
The coefficients $C_{s_{1}s_{2}s_{3}}^{s}$ follow from the relation
\[
\phi _{s_{1}}({\bf n})\phi _{s_{2}}({\bf n})\phi _{s_{3}}({\bf n}%
)=\sum_{s}C_{s_{1}s_{2}s_{3}}^{s}\phi _{s}({\bf n}). 
\]

For temperatures below $T_{c}$, satisfying the inequality $%
(T_{c}-T)/T_{c}\ll 1$, Eq.(\ref{12}) can be rewritten as
\begin{equation}
\sum_{s=0}^{\infty }\Delta _{s}\left( \frac{\lambda _{0}-\lambda _{s}}{%
\lambda _{0}}+\lambda _{s}\frac{T_{c}-T}{T_{c}}\right) \phi _{s}({\bf n})-%
\frac{7\zeta (3)}{8\pi ^{2}T_{c}^{2}}\sum_{s=0}^{\infty }\lambda _{s}\phi
_{s}({\bf n})\left( \sum_{\left\{ s_{i}\right\}
}C_{s_{1}s_{2}s_{3}}^{s}\Delta _{s_{1}}\Delta _{s_{2}}\Delta _{s_{3}}\right)
=0,  \label{13}
\end{equation}
where we neglect higher powers of $(T_{c}-T)/T_{c}$. It can be easily seen
from Eq.(\ref{13}) that for $T\rightarrow T_{c}$ one has $\Delta _{0}\sim
(T_{c}-T)^{1/2}$, and that $\Delta _{s}$ with $s>0$ are either
equal to zero or proportional to $%
 (T_{c}-T)^{3/2}$. Therefore, the
equation for the leading
coefficient $\Delta _{0}$ is:
\[
\frac{T_{c}-T}{T_{c}}\Delta _{0}-\frac{7\zeta (3)}{8\pi ^{2}T_{c}^{2}}%
C_{000}^{0}\Delta _{0}^{3}=0,
\]
where the coefficient $C_{000}^{0}$ is equal to $3/2$. We thus obtain the
following expression for the order parameter on the Fermi surface ($\xi =0$%
): 
\begin{equation}
\Delta (0,{\bf n})=\frac{4\pi }{\sqrt{21\zeta (3)}}T_{c}\sqrt{\frac{T_{c}-T}{%
T_{c}}}\phi _{0}({\bf n})=2.5\,T_{c}\sqrt{\frac{T_{c}-T}{T_{c}}}\phi _{0}(%
{\bf n});\quad \frac{T_{c}-T}{T_{c}}\ll 1.  \label{14}
\end{equation}

For $\xi \neq 0$, i.e. $p\neq p_{F}$, the order parameter can be calculated by
using Eq.(\ref{8}). Fig.2 shows the numerically calculated dependence of
the order parameter on the
modulus of the momentum ${\bf p}$ for various
values of the angle $\theta $ between the vector $ {\bf p}$ and the direction
of dipoles.
Note that for both $s$ and $p$-wave pairing due to a short-range
interaction, the order parameter is momentum independent for
$p$
satisfying the condition of the ultra-cold limit and rapidly decays at
larger
$p$. The momentum dependence of the order parameter for dipolar gases
results in a nonuniform energy gap for single-particle excitations and can,
for example, manifest itself in processes with a large (of the order of $p_F$) 
momentum transfer.
\begin{figure}[tbp]
\epsfxsize10cm 
\centerline{\epsfbox{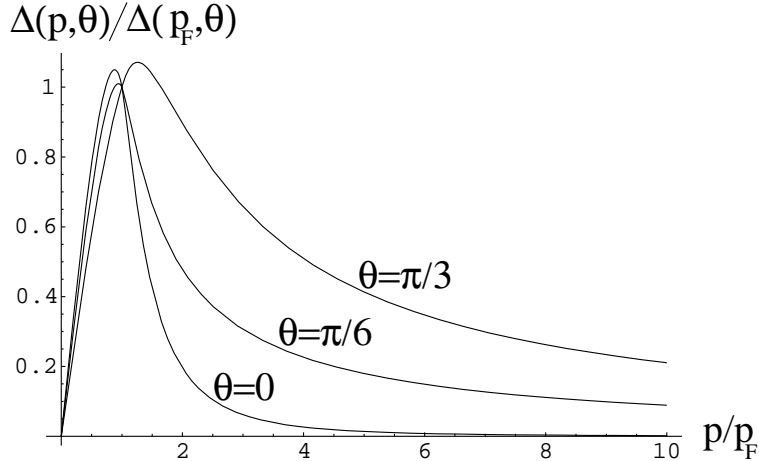}}
\caption{The order parameter $\Delta(p,\theta)$ (in units of
$\Delta(p_F,\theta)$) as a function of the momentum $p$ (in units of $p_F$)
for various values of the polar angle $\theta$.}
\end{figure}

In Eq.(\ref{14}) the anisotropy of the order parameter in the momentum space
is described by the function $\phi_0({\bf
n})=\sqrt{2}\sin{[(\pi/2)\cos\theta]}$.  The order parameter is an odd
function of $\cos\theta$ and is negative for $ \pi/2<\theta\leq\pi$. This
does not cause any problems as all physical quantities are determined by
$|\Delta|$. The maximum value of $|\Delta(0, {\bf n})|$ is reached in the
direction of the dipoles, i.e. for $\theta=0$ ($\phi_0=\sqrt{2}$). In the
direction perpendicular to the dipoles ($\theta=\pi/2$) the order parameter
vanishes.

If we consider only the $p-p$ scattering channel, the angular dependence
of the order parameter will be determined by the function $\sqrt{3}\cos\theta
$ instead of $\phi _{0}({\bf n})$. The coefficient $C_{000}^{0}$ is then
equal to $9/4$, and the result for the order parameter reads: 
\begin{equation}
\Delta _{pp}(0,{\bf n})=\frac{4\pi }{3}\sqrt{\frac{2}{7\zeta (3)}}\,T_{c} 
\sqrt{\frac{T_{c}-T}{T_{c}}}\sqrt{3}\cos\theta=2.5\,T_{c}\sqrt{\frac{
T_{c}-T} {T_{c}}}\sqrt{2}\cos\theta;\quad \frac{T_{c}-T}{T_{c}}\ll 1.
\label{14.1}
\end{equation}
The angular dependence of $|\Delta_{pp}(0,{\bf n})|$ is
qualitatively
similar to that of the true order parameter $|\Delta(0,{\bf n})|$. The
maximum value of $|\Delta_{pp}(0,{\bf n})|$ is also reached in the direction
of the dipoles and it is exactly equal to the maximum value of $|\Delta(0,
{\bf n})|$. Also, $|\Delta_{pp}(0,{\bf n})|$ vanishes in the perpendicular
direction. However, for intermediate values of $\theta$ the quantity $
|\Delta_{pp}(0,{\bf n})|$ can be up to $40\%$ smaller than $|\Delta(0,{\bf n}
)|$.

The anisotropy of the order parameter provides a major difference of the
properties of the superfluid dipolar Fermi gas from those of the
(two-component) fermionic gas with the $s$-wave pairing due to short-range
intercomponent interaction. This anisotropy ensures the anisotropic momentum
dependence of the gap in the spectrum of single-particle excitations, which
appears below the transition temperature $T_c$. For example, excitations
with momenta in the direction of the dipoles acquire the largest gap. In
contrast to this, the eigenenergies of excitations with momenta
perpendicular to the
 dipoles remain unchanged. The properties of collective
excitations are also
 expected to have a nontrivial dependence on the
direction of their momenta.
 Therefore, the response of the dipolar superfluid
Fermi gas to small
 external perturbations will have a pronounced anisotropic
character.

Another distinguished feature of the superfluid dipolar Fermi gas is related
to the temperature dependence of the specific heat. Well below the critical
temperature the single-particle contribution to the specific heat is
proportional to $T^{2}$, rather than being exponentially small as in the
case of the $s$-wave pairing. This follows from the fact that the energy 
$\varepsilon$ of single-particle excitations has a line of zeros on the 
Fermi surface: $\varepsilon(p_F)=0$ for the angles at which
$\Delta(p_F,{\bf n})=0$, i.e. for $\theta=\pi/2$ and an arbitrary azimuthal
angle
 $\varphi$. As a consequence, the density of states in
the vicinity of the Fermi energy is $ \nu (\varepsilon )\sim \varepsilon $
for $\varepsilon \ll \Delta _{0}$. Therefore, at temperatures $T\ll \Delta
_{0}\sim T_{c}$, the temperature dependent part of the energy of the system
is proportional to $T^{3}$, and the specific heat is hence proportional to
$T^{2}$. This contribution is much larger than the one of collective modes
which is $\propto T^{3}$ and is dominant in the case of the $s$-wave pairing.

It should also be mentioned that the properties of the superfluid dipolar
fermionic gas are different from the properties of the gas with the $p$-wave
pairing originating from a short-range attractive interaction in the $p$
-wave channel. The reason is that in the latter case the order parameter is
isotropic, similarly to the $B$-phase of superfluid $^{3}$He. The order
parameter of dipolar gases is also different from the order parameter of the
$A$ phase of $^{3}$He where the gap vanishes only at two points on the Fermi
sphere, i.e. at the poles of the sphere $\theta =0$ and $\theta=\pi $. 

The anisotropy of the order parameter of dipolar gases is similar to that in
the polar phase of superfluid liquid $^{3}$He, not realized in
experiments as it has higher energy than experimentally observed $A$
and $B$ phases (see, e.g. Ref. \cite{3He}). For the polar phase the order 
parameter is also equal to zero on the equator of the Fermi sphere 
($\theta=\pi/2$ and an arbitrary $\varphi$). The situation where the order
parameter is zero on one or several lines on the Fermi surface is encountered
in heavy-fermion compounds (for a review of possible superconducting
phases of heavy-fermion compounds belonging to different crystalline groups
see, e.g. Ref. \cite{heavy-fermions}). In these cases the temperature
dependence of the specific heat is also $\propto T^{2}$ (see, e.g. 
Ref. \cite{T2}. However, from a
general point of view, one would expect a different physical behavior of
dipolar gases, for example with regard to the frequency and angular
dependence of the response. This is because of the different types of 
symmetry groups: continuous rotational group for dipolar gases and discrete 
crystalline group in the case of heavy-fermion materials
(see Ref. \cite{heavy-fermions} for more details).  

\section{Concluding remarks}

Our results show prospects for achieving the BCS transition in
single-component trapped gases of dipolar particles, in particular for
(electrically polarized) fermionic polar molecules. As has  been shown in
Refs. \cite{dipstoof,barpetr,castin}, for trapped gases the BCS  transition
temperature is close to that of the 3D uniform gas with density $n$ equal to
the maximum density in the trap. This requires $T_c$ to be much  larger than
the maximum trap frequency, which is generally the case for  achievable
temperatures. Therefore, we will use Eq.(\ref{11}) for estimating $T_c$ in
the trapped case.

We first compare our equations (\ref{TBCS}) and (\ref{11}) with the
well-known BCS formula (see e.g. \cite{LL9}) for the two-component Fermi gas
with short-range attractive intercomponent interaction. In the latter case
the exponent is expressed in terms of the $s$-wave scattering length $a$ and
is equal to $\pi\hbar /2p_F \left| a \right|$. We then see that our
dipole-dipole scattering with odd orbital angular momenta is equivalent to
having the $s$-wave scattering length 
\begin{equation}
a_d = - \frac{2 m d^{2}}{\pi^2 \hbar^2}.  \label{scattlength}
\end{equation}
Accordingly, Eq.(\ref{11}) takes the form 
\begin{equation}
T_c = 1.44 \varepsilon_F \exp\left\{- \frac{\pi \hbar}{2p_{F}\left| a_d
\right|}\right\}.  \label{sTc}
\end{equation}
Qualitatively, the result of Eq.(\ref{scattlength}) is more or less
expected, since $\left| a_d \right|$ turns out to be of the order of the
characteristic radius of the dipole-dipole interaction, $r_{*}\sim md^2
/\hbar^2$, introduced in Section III.

For most polar molecules the electric dipole moment ranges from $0.1$ to 
$1$ Debye. For example, the dipole moment of fermionic ammonia
molecules $^{15}$ND$_3$ \footnote{
Bosonic ($^{14}$ND$_3$) and fermionic ($^{15}$ND$_3$) ammonia molecules
were recently trapped at temperatures of around $30$ mK in the
Rijnhuizen  experiment \cite{movefield}.} 
is $d=1.5$ D, and we have the effective scattering length is $a_d = -1450$
\AA.
 This is larger than the scattering length for the intercomponent
interaction in the widely discussed case of two-species fermionic gas of $%
^{6}$Li. From Eq.(\ref{sTc}) we find that the BCS transition temperature for
the single-component ND$_3$ dipolar gas will be larger than $100$ nK at
densities $n>5\cdot 10^{12}$ cm$^{-3}$. Another interesting example is 
a linear fermionic molecule HCN which has dipole moment $d=2.98$ D, and the
corresponding effective scattering length $a_d=-7400$ \AA.

Remarkably, in ultra-cold single-component fermionic gases one can hope to
reach much higher densities than in Bose gases and think of achieving the
BCS transition for a smaller effective scattering length $a_d$. The reason
is that inelastic decay processes will be strongly suppressed due to the
Pauli exclusion principle. For two identical fermions with momentum $p$
of the relative motion, the pair correlation function behaves as $(pr/\hbar)^2$
at interparticle distances $r$ smaller
 than the de Broglie wavelength
$\hbar/p$. Generally, inelastic processes
 occur at short interparticle
distances $R_0$ which in the ultra-cold limit
 are much smaller than
$\hbar/p$. Therefore, two-body inelastic collisions will be suppressed as
$(pR_0)^2$ compared to the bosonic case where the pair correlation function
is of order unity at any $r$ outside the region of interparticle interaction.
As a result, in a non-degenerate gas of fermions the inelastic rate decreases 
with temperature and is suppressed as $ T/\varepsilon_0$, where the energy
$\varepsilon_0=\hbar^2/mR_0^2$. In a
 quantum degenerate Fermi gas a
characteristic momentum of particles is of
 the order of $p_F$ and the
suppression factor is $\sim
 (\varepsilon_F/\varepsilon_0)$. The suppression
of two-body inelastic collisions in fermionic gases was first found for spin
relaxation in atomic deuterium \cite{Verhaar}. For the rate of 3-body
recombination we expect even a stronger suppression, i.e. by a factor of
$(T/\varepsilon_0)^2$ for the non-degenerate gas, and by a factor of
$(\varepsilon_F/\varepsilon_0)^2$ in the regime of quantum degeneracy
($T<\varepsilon_F$). The suppression factor for inelastic collisions in
single-component quantum degenerate Fermi gases,
$(\varepsilon_F/\varepsilon_0)$, is of the order of $ (nR_0^3)^{2/3}$. The
distance $R_0$ is commonly smaller than $50$ \AA. Therefore, even at
densities $n\sim 10^{16}$ cm$^{-3}$ one expects the suppression of two-body
inelastic rates at least by 2 orders of magnitude, and the suppression of
3-body recombination by 4 orders of magnitude. This allows us to think of
achieving densities $n\sim 10^{16}$ cm$^{-3}$ or
somewhat larger, which is by
more than a factor of $10$ higher than the densities currently reached with
ultra-cold Bose gases.

For $n\sim 10^{16}$ cm$^{-3}$, the BCS transition temperature $T_{c}$ is in
the nanokelvin or microkelvin regime for dipolar fermionic gases with an
effective scattering length $a_{d}$ ranging from $-20$ to $-30$ \AA . Such
values of $a_{d}$ one easily finds in molecular gases. For example,
fermionic $^{14}$N$^{16}$O molecules have dipole moment $d=0.16$ D and Eq.(%
\ref{scattlength}) gives $a_{d}\approx -24$ \AA .

Interestingly, the effective scattering length $|a_{d}|$ of the order of
several tens of angstroms can be obtained in gases of atoms with induced
dipole moments \cite{magnetic,CO}. By using a high dc electric field ($\sim
10^{6}$ V/cm) \cite{you-marinescu0} one can induce permanent atomic dipole
moments close to $0.1D$. The same values of $d$ one obtains for the time
averaged dipole moment of an atom, induced by a stroboscopic laser coupling
of the ground atomic state to a Rydberg state \cite{rydberg}. The
corresponding scattering length $a_{d}$ can then be close to $-30$ \AA , and
one can think of achieving the BCS transition in such single-component
atomic dipolar gases at densities $n\sim 10^{16}$ cm$^{-3}$ and temperatures 
$\sim 100$ nK.

\section*{Acknowledgments}

We acknowledge fruitful discussion with M. Lewenstein and. G. Meijer.
This work was
financially supported by the Nederlandse Organisatie voor Wetenschappelijk
Onderzoek (NWO), by the Stichting voor Fundamenteel Onderzoek der Materie
(FOM), by Deutsche Forschungsgemeinschaft (DFG), by the Alexander von
Humboldt Stiftung, and by the Russian Foundation for Basic Research.


\begin{references}
\bibitem{orbital}  This has been first pointed out in the studies of spin
relaxation collisions \cite{Kagan,Verhaar}, and was then expounded in \cite
{you-marinescu0,you-marinescu}.
\bibitem{magnetic}  In the case of atoms with large magnetic moments one can
also think of efficient elastic scattering due to the magnetic dipole
interaction. For chromium atoms, the magnetic moment is $6\mu _{B}$, where $%
\mu _{B}$ is the Bohr magneton. This is equivalent to having the dipole
moment $d=0.056$ D, and Eq.(\ref{scattlength}) leads to $a_{d}\approx -5$ 
\AA . Eq.(\ref{sTc}) then shows that even at nanokelvin temperatures the BCS
transition requires extremely high densities at which the gas will be
unstable because of recombination processes.
\bibitem{CO} The situation is likely to be
same for fermionic $^{13}$C$^{16}$O molecules in the ground electronic
state, which have dipole moment $d=0.1$ D and effective scattering length 
$a_{d}\approx -9$ \AA . On the other hand, the dipole moment of the metastable molecule
$^{13}$C$^{16}$O in the first electronically excited $a^3 \Pi$ state 
is equal to $1.37$D, and the effective scattering length is 
$a_{d} \approx -1670$ \AA . This opens interesting possibilities to
(optically) manipulate the permanent dipole moment $d$ and the scattering
length $a_d$ in CO molecular gases.
\end{references}
\end{document}